# Structural Transition of Li$_2$RuO$_3$ Induced by Molecular-Orbit Formation

Yoko Miura, Masatoshi Sato,* Youichi Yamakawa,[1] Tatsuro Habaguchi,[1] and Yoshiaki Ōno[1]

*Department of Physics, Nagoya University, Furo-cho, Chikusa-ku, Nagoya 464-8602, Japan*

[1]*Department of Physics, Niigata University, Ikarashi, Nishi-ku, Niigata 950-2181, Japan*

**Abstract**

A pseudo honeycomb system Li$_2$RuO$_3$ exhibits a second-order-like transition at temperature $T=T_c \sim 540$ K to a low-$T$ nonmagnetic phase with a significant lattice distortion forming Ru-Ru pairs. For this system, we have calculated the band structure, using the generalized gradient approximation (GGA) in both the high- and low-$T$ phases, and found that the results of the calculation can naturally explain the insulating behavior observed in the low-$T$ phase. The detailed characters of the Ru 4$d$ $t_{2g}$ bands obtained by the tight-binding fit to the calculated dispersion curves show clear evidence that the structural transition is driven by the formation of the Ru-Ru molecular-orbit, as proposed in our previous experimental studies.

*Corresponding author (M. Sato): e43247a@nucc.cc.nagoya-u.ac.jp

## 1. Introduction

Systems with the so-called honeycomb structure are characterized by the two-dimensionality and the small coordination number of 3, and exhibit a variety of physical properties arising from these characteristics. For example, their magnetic characters are often expected to be unusual.[1] In Na$_3$Cu$_2$SbO$_6$, due to the Jahn-Teller distortion induced by Cu$^{2+}$ ions, the effective coordination number is reduced to only 2, and the Cu$^{2+}$ spins can be described by the ferromagnetic-antiferromagnetic one-dimensional alternating chain model, where spin-gap behavior has been found.[2,3]

Among conducting systems with the honeycomb structure, Li$_x$ZrNCl[4,5] and Li$_x$(THF)$_y$HfNCl[6] (THF: tetrahydrofuran) are known to be superconducting, and the transition temperature of the latter is as high as $\sim$ 25.5 K. Li$_2$RuO$_3$ with pseudo honeycomb structure exhibits a structural transition, at $\sim$540 K, to a nonmagnetic insulating state with zero spin susceptibility.[7] The transition is accompanied by a significant lattice distortion to form Ru-Ru pairs. Based on these experimental results, we have proposed that the transition is driven by the formation of molecular-orbits of Ru-Ru pairs, which can be considered to be new type in the sense that it is a second order (or, at least, a nearly second order) transition occurring with varying $T$.

It is interesting to study this transition further and to compare it with those of other Ru oxides, because there are several systems which exhibit similar transport and magnetic anomalies associated with their structural transitions. La$_4$Ru$_2$O$_{10}$[8-10] and Tl$_2$Ru$_2$O$_7$[11,12] are the examples of such systems.

The present work has been carried out to investigate the mechanism of the structural transition of Li$_2$RuO$_3$ further, where the band structure calculations followed by the tight binding fits have been done to examine the

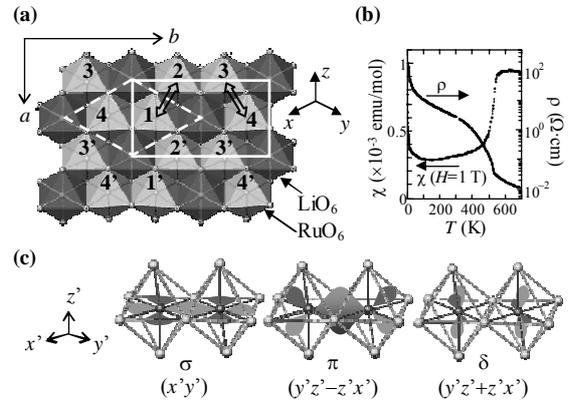

Fig. 1 (a) Schematic figure of the Li$_2$RuO$_3$ honeycomb layer viewed from the direction perpendicular to the *ab*-plane. O atoms are at the corners of the octahedra and Ru or Li atom is within each octahedron. Li atoms are between the honeycomb layers. Ru sites 1', 2', 3' and 4' are equivalent of those of 1, 2, 3 and 4, respectively. In the high-$T$ phase, 1 and 2 are equivalent to 3 and 4, respectively. The $x$-, $y$- and $z$- axes are defined for Ru 4$d$ and O 2$p$ orbits. The open arrows indicate the Ru-Ru bonds whose lengths undergo the significant reduction through the phase transition. The rectangle shown by the solid lines is the unit cell for the low-$T$ phase and the rhombus shown by the broken line indicates the unit cell of the high-$T$ phase. (b) Magnetic susceptibility $\chi$ measured under magnetic field $H=1$ T and the electrical resistivity $\rho$ of Li$_2$RuO$_3$ are shown against $T$. (c) Schematic figures of the wave functions of $\sigma$-, $\pi$- and $\delta$-molecular orbits (bonding orbits) formed by the pairs of Ru 4$d$ $t_{2g}$ orbits ($x'y'$, ($y'z'-z'x'$), ($y'z'+z'x'$)). The black and gray circles are Ru and O atoms, respectively.

Table 1  The structural parameters obtained by the powder neutron Rietveld analysis at 600 K ($>T_c$) are shown.

| Atom | Site | $x$ | $y$ | $z$ | B (Å$^2$) |
|---|---|---|---|---|---|
| Ru | 4g | 0 | 0.3308(3) | 0 | 1.65(2) |
| Li1 | 2a | 0 | 0 | 0 | 2.68(6) |
| Li2 | 4h | 0 | 0.3425(9) | 1/2 | 2.68(6) |
| Li3 | 2c | 0 | 0 | 1/2 | 2.68(6) |
| O1 | 8j | -0.0162(7) | 0.1701(2) | 0.2324(2) | 1.36(1) |
| O2 | 4i | 0.4991(8) | 0 | 0.2325(4) | 1.36(1) |

Space group $C2/m$; $a$=5.0466(3) Å, $b$=8.7649(2) Å, $c$=5.9417(3) Å, $\beta$=124.495(4)°, $R_{wp}$=4.32, $R_e$=3.07, $S$=1.41

Table 2  The structural parameters obtained by the powder neutron Rietveld analysis at room temperature ($<T_c$) are shown.

| Atom | Site | $x$ | $y$ | $z$ | B (Å$^2$) |
|---|---|---|---|---|---|
| Ru | 4f | 0.2737(6) | 0.0766(2) | -0.0063(6) | 1.05(2) |
| Li1 | 2e | 0.706(3) | 1/4 | -0.068(2) | 1.11(5) |
| Li2 | 4f | 0.253(3) | 0.0991(7) | 0.493(3) | 1.11(5) |
| Li3 | 2e | 0.772(4) | 1/4 | 0.513(5) | 1.11(5) |
| O1 | 4f | 0.7553(7) | 0.0805(6) | 0.2489(6) | 0.61(1) |
| O2 | 4f | 0.7757(6) | 0.0819(6) | 0.7685(6) | 0.61(1) |
| O3 | 2e | 0.255(1) | 1/4 | 0.2144(9) | 0.61(1) |
| O4 | 2e | 0.267(1) | 1/4 | 0.761(1) | 0.61(1) |

Space group $P2_1/m$; $a$=4.9210(2) Å, $b$=8.7829(2) Å, $c$=5.8941(2) Å, $\beta$=124.342(2)°, $R_{wp}$=4.17, $R_e$=3.07, $S$=1.36

orbital characters of the Ru 4d $t_{2g}$ bands. We have found rather clear evidence that the molecular-orbit formation works as the driving mechanism of the transition.

## 2. Structural Transition of Li$_2$RuO$_3$

In this section, we briefly describe the characteristics of the phase transition of Li$_2$RuO$_3$.[7] It has Ru$_2$LiO$_6$ honeycomb layers, which consist of the edge-sharing RuO$_6$- and LiO$_6$-octahedra (see Fig. 1(a)) with Li layers between them. The structural transition was found at $T$=$T_c$~540 K. With decreasing $T$, the electrical resistivity $\rho$ increases anomalously at $T_c$, and the magnetic susceptibility $\chi$ begins to decrease sharply as shown in Fig. 1(b). No hysteresis was observed in our experimental accuracy, indicating that the transition is second order at least nearly second order. The space group is $C2/m$ above $T_c$ and $P2_1/m$ below $T_c$. Tables 1 and 2 show the parameters determined by the neutron Rietveld analyses above and below $T_c$, respectively. Above $T_c$, the Ru atoms have a slight distortion from the ideal honeycomb structure, and upon the structural transition, two of the six bond lengths between neighboring Ru atoms within a hexagon exhibits a significant reduction. The bond lengths indicated by the arrows in Fig. 1(a) are shorter by ~19 % than the others within a hexagon at room temperature (RT). The atomic distances between the Ru1 and Ru2 (1-2), 3-4, 1'-2' and 3'-4' are shorter than the other distances. Even in the group with longer distances, the distances 4-1, 2-3, 4'-1' and 2'-3' are shorter by ~0.1 % than 1-2' and 3'-4. (Note that the Ru sites labeled by the numbers 1'~4' are crystallographically equivalent to 1~4, respectively.)

In the high-$T$ phase, the distances 1-2, 1-2', 1'-2', 3-4, 3'-4 and 3'-4' are shorter than those of 4-1, 2-3, 4'-1' and 2'-3' by ~2.5 %. (We use the labels 1~4 and 1'~4' even above $T_c$, although there are only two crystallographically distinct Ru sites. The Ru1 and Ru2 are equivalent to Ru3 and Ru4, respectively.) The rhombus and rectangle indicate the primitive unit cells above and below $T_c$, respectively.

We have proposed,[7] considering the very drastic change of the Ru-Ru bond lengths, the molecular-orbit formation is the primary driving force of the transition: The σ-, π-, δ-bonding and σ$^*$-, π$^*$- and δ$^*$-antibonding molecular orbits of the Ru-Ru atoms are formed below $T_c$. (For the bonding orbits, see Fig. 1(c), where the wave functions of the three molecular orbits of 4d $t_{2g}$ orbits are shown by using the local coordinates $x'$, $y'$ and $z'$. These orbits are expected to order σ-, π-, δ-, δ$^*$-, π$^*$- and σ$^*$-orbits from the lower energy side.) The eight 4d $t_{2g}$ electrons within the Ru$^{4+}$-Ru$^{4+}$ pair atoms occupy the σ-, π-, δ- and δ$^*$-orbits, resulting in the nonmagnetic and insulating state. The large change of the bond lengths as well as the changes of the transport and magnetic behavior observed through the transition can be naturally understood by this picture.

## 3. Band Calculations and the Tight Binding Fits

The density functional calculations have been performed by using the generalized gradient approximation (GGA),[13] where we have used the WIEN2k package.[14] The structural parameters[7] shown in Tables 1 and 2 are used in the calculations above and below $T_c$, respectively.

The calculated density of states (DOS) near the Fermi level $E_f$ (=0 eV) is shown in Fig. 2, both in the high-$T$ (dotted line) and low-$T$ (solid line) phases. The data indicate that the high-$T$ phase is metallic. From the experimental results of the resistivity shown in Fig. 1(b), we could not clearly distinguish if the high-$T$ phase was metallic or not, because the negative value of the temperature derivative of the resistivity, d$\rho$/d$T$ observed just above $T_c$ may be due to the effect of the fluctuation near the transition. In the low-$T$ phase, the existence of an energy gap is consistent with the transport measurements. Although DOS was reported at RT in a recent paper,[15] it was derived for the structure reported in an earlier work with the space group $C2/c$.[16]

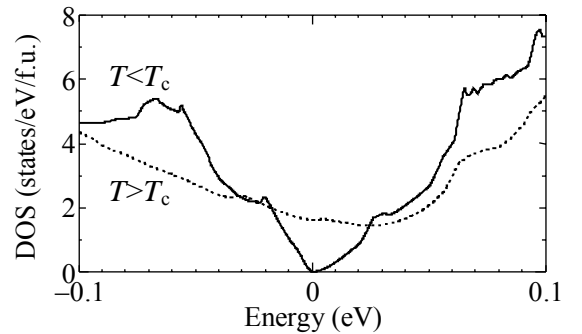

Fig. 2  Calculated density of states (DOS) per formula unit of Li$_2$RuO$_3$ near the Fermi level both of the high-$T$ (dotted line) and low-$T$ (solid line) phases. The Fermi level $E_f$ is at 0 eV.

To extract the orbital characters of the bands calculated by GGA, the bands dispersion obtained by the tight-binding model are fitted to those of GGA. In the tight-binding calculations, we have used the two-dimensional $d$-$d$ model with the nearest neighbor electron transfers among the Ru $4d$ $t_{2g}$ ($xy$, $yz$, $zx$) orbits and/or $d$-$p$ model with the nearest neighbor electron transfers among the Ru $4d$ $t_{2g}$ ($xy$, $yz$, $zx$), $e_g$ ($x^2$-$y^2$, $3z^2$-$r^2$) and O $2p$ ($x$, $y$, $z$) orbits, where the $x$, $y$ and $z$ axes are defined as shown in Fig. 1(a).

In the high-$T$ phase, the 28 band $d$-$p$ model has been found to give satisfactory results, while the 6 band $d$-$d$ model has not presented satisfactory results. In the low-$T$ phase, the 12 band $d$-$d$ model have presented satisfactory results, while we have not found a firm result by the 56 band $d$-$p$ model, possibly because the number of the parameters is too large.

The left and right panels of Fig. 3 show the points of the reciprocal space in the high-$T$ ($z$=2) and low-$T$ ($z$=4) phases, respectively. In the high-$T$ phase, Γ and Γ' are equivalent, but M and M' are not equivalent. In the low-$T$ phase, Γ, Σ, Y and X correspond to (0, 0), ($\pi/a_L$, $\pi/b_L$), (0, $\pi/b_L$) ($\pi/a_L$, 0) points in the reciprocal space. The $a_L$ and $b_L$ are the lattice parameters in the low-$T$ phase.

In Figs. 4 and 5, the GGA band structures around the Fermi level $E_f$ (=0 eV) are shown by the solid lines for high-$T$ and low-$T$ phases, respectively. The results of the tight-binding fits are also shown by the broken lines. Their overall features can be described as follows. All bands shown here are mainly characterized by the contribution from Ru $4d$ $t_{2g}$ ($xy$, $yz$, $zx$) orbitals, and the other bands are well apart from these bands. In particular, the Li bands are far away from the Fermi level.

Now, we first show the results of high-$T$ phase in more detail. The $4d$ $t_{2g}$ bands obtained by GGA (solid lines) are located between -1.0 and 0.4 eV. They are well separated from the O $2p$ bands located between -6.9 and -2.2 eV and from the Ru $e_g$ bands between 2.6 and 3.4 eV. In Fig. 4, the dispersion curves of the Ru $4d$ $t_{2g}$ bands are shown, where we can see that several bands cross the Fermi level. We can also see that the tight-binding fits (dotted lines) carried out by using the 28 band $d$-$p$ model can qualitatively reproduce the GGA results. In Table 3, the weights of Ru1 $4d$ $t_{2g}$ orbitals obtained by the fittings are shown at Γ and M points. (Note that the $yz$- and $zx$-orbitals are crystallographically equivalent.) The weights of Ru2 4d $t_{2g}$ orbits are equal to those of Ru1.

Next, we show the results for the low-$T$ phase. The energy region of the Ru $4d$ $t_{2g}$ bands is broadened by a

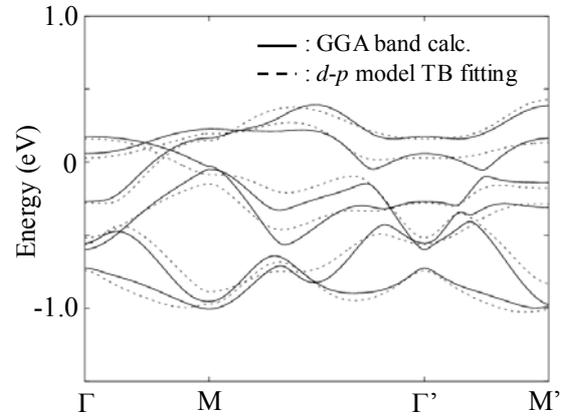

Fig. 4 Band structure of the high-$T$ phase of Li$_2$RuO$_3$ is shown near the Fermi level. The solid lines have been obtained by GGA and the dotted lines are the results of the tight-binding fits by the 28 band $d$-$p$ model. The Fermi level $E_f$ is at 0 eV.

Table 3 The weights of Ru1 $4d$ $t_{2g}$ ($xy$, $yz$, $zx$) orbits at 600 K (> $T_c$) are shown per one Ru atom at Γ and M points for the bands with various energies. The 28 band $d$-$p$ model was used to fit the results of the nearest-neighbor tight-binding calculation to the GGA results. The weights of Ru2 are equal to those of Ru1.

Γ point

| energy of the tight-binding calculation (eV) | $xy$ | $yz$ | $zx$ |
|---|---|---|---|
| 0.16 | 0.000 | 0.183 | 0.183 |
| 0.03 | 0.156 | 0.104 | 0.104 |
| -0.28 | 0.134 | 0.124 | 0.124 |
| -0.52 | 0.000 | 0.179 | 0.179 |
| -0.56 | 0.230 | 0.062 | 0.062 |
| -0.74 | 0.211 | 0.095 | 0.095 |

M point

| energy of the tight-binding calculation (eV) | $xy$ | $yz$ | $zx$ |
|---|---|---|---|
| 0.19 | 0.332 | 0.015 | 0.015 |
| 0.15 | 0.020 | 0.190 | 0.190 |
| -0.08 | 0.000 | 0.192 | 0.192 |
| -0.15 | 0.000 | 0.164 | 0.164 |
| -0.89 | 0.313 | 0.008 | 0.008 |
| -0.97 | 0.053 | 0.178 | 0.178 |

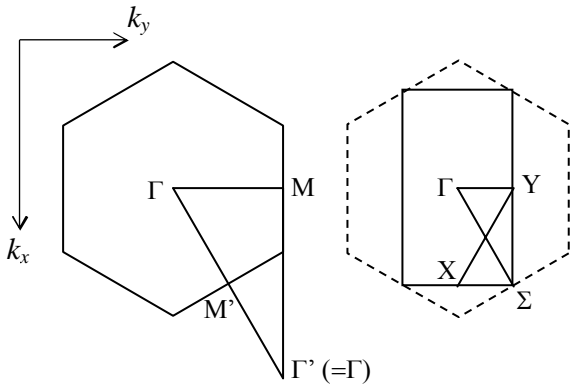

Fig. 3 Various points in the reciprocal spaces of the honeycomb layers are shown. The left and right panels are for the high-$T$ and low-$T$ phases, respectively.

factor of ~ 2 as compared with that above $T_c$, while the dispersion of each band becomes weaker. The fitted lines (dotted lines) obtained by the 12 band $d$-$d$ model can qualitatively reproduced the GGA results (solid lines).

Here, in order to make it easy to construct the molecular orbit picture, we change the bases of the $t_{2g}$ wave functions to $x'y'$, $(y'z'-z'x')$ and $(y'z'+z'x')$ orbits defined with the local coordinates $x'$, $y'$ and $z'$. As shown in Fig. 1(c), the molecular orbits $\sigma$, $\pi$ and $\delta$ are constructed by the overlapping of the two neighboring orbits with the same characters.

In Table 4, we show the weights of these Ru1 $4d$ $t_{2g}$ orbits at $\Sigma$ and Y points, where the two fold degeneracy exists. The weights of Ru2, Ru3 and Ru4 sites are equal to those of Ru1. The top and bottom bands have large weights of $x'y'$ orbital, indicating that they correspond to the $\sigma^*$ (antibonding) and $\sigma$ (bonding) molecular orbits, respectively. The second and third bands from the top and bottom have the large weights of $(y'z'-z'x')$ and $(y'z'+z'x')$ orbits, and can be considered to the $\pi^*$ and $\delta^*$ (antibonding) and $\pi$ and $\delta$ (bonding) molecular orbits, respectively. The electron transfer energy $t_\sigma$ between the neighboring $x'y'$ orbits has been found to be 1.06 eV for the Ru-Ru pairs with the shorter bond length. The transfer energies $t_\pi$ and $t_\delta$ between the $(y'z'-z'x')$ orbits and $(y'z'+z'x')$ orbits are 0.37 and 0.14 eV, respectively, between the Ru-Ru atoms with the shorter bond length. The transfers between Ru and Ru of the longer spacing are much smaller than that of the shorter one.

## 4. Discussion

By the above analyses, we have found that in the high-$T$ phase, we do not see any characteristic distribution of the weights of the three Ru $4d$ $t_{2g}$ orbits to indicate a certain type ordering. In contrast, below $T_c$, it is clear that the electron bands split into the bonding ($\sigma$, $\pi$ and $\delta$) and antibonding ($\sigma^*$, $\pi^*$ and $\delta^*$) ones, where the splitting between $\sigma$ and $\sigma^*$ is the largest and it is the smallest between $\delta$ and $\delta^*$. The result indicates the molecular orbit formation below $T_c$.

Above $T_c$, we had to adopt the 28 band $d$-$p$ model, instead of the effective $d$-$d$ model in the tight-binding fits to reproduce the dispersion curves obtained by GGA. It is probably because the consideration of only the nearest neighbor electron transfer is not satisfactory for the undistorted conducting phase. In contrast, the 12 band effective $d$-$d$ model has been found to be able to reproduce the GGA results below $T_c$. It is understood by considering the localized nature of the electrons in the low-$T$ phase.

Jackeli and Khomskii[17] proposed a transition mechanism of the present system. They treated the present system as a Mott insulator and deduced the formation of Ru-Ru singlet dimers. In contrast to their arguments we have shown the transition of $Li_2RuO_3$ by the approach from the view that it is metallic above $T_c$.

As stated in section 1, there are several compounds which have structural transitions accompanied with the significant anomalies of the electrical resistivity $\rho$ and the magnetic susceptibility $\chi$ similar to those of the present system. $La_4Ru_2O_{10}$[8-10] and $Tl_2Ru_2O_7$[11,12] are the examples of such systems, although their transitions are first order in contrast to that of $Li_2RuO_3$. In the former case, the localize spins $S=1$ exist in the high-$T$

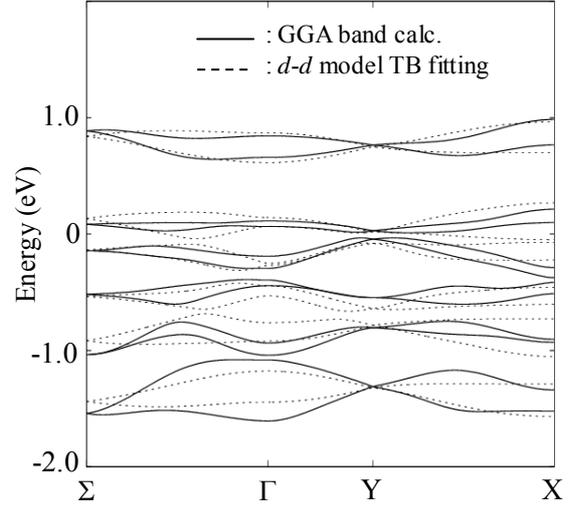

Fig. 5  Band structure of the low-$T$ phase of $Li_2RuO_3$ near the Fermi level in the low-$T$ phase. The solid lines are obtained by GGA and the dotted lines are the results of the tight-binding fits by the 12 band $d$-$d$ model. The Fermi level $E_f$ is at 0 eV.

Table 4  The weights of Ru1 $4d$ $t_{2g}$ ($xy$, $yz$, $zx$) orbits at room temperature ($< T_c$) are shown per one Ru atom at $\Sigma$ and Y points for the bands with various energies. At each energy, two bands are degenerated. The weights were estimated by the nearest-neighbor tight-binding fits, using the 12 band $d$-$d$ model. The bases are changed from $xy$, $yz$ and $zx$ to $x'y'$, $(y'z'-z'x')$ and $(y'z'+z'x')$ orbits (see Fig. 1(a) and (c) for the axes). The values for Ru2, Ru3 and Ru4 are equal to those for Ru1.

$\Sigma$ point

| energy of the tight-binding calculation (eV) | $x'y'$ | $y'z'-z'x'$ | $y'z'+z'x'$ |
| --- | --- | --- | --- |
| 0.84 | 0.495 | 0.003 | 0.003 |
| 0.13 | 0.002 | 0.476 | 0.022 |
| -0.14 | 0.005 | 0.017 | 0.478 |
| -0.54 | 0.013 | 0.020 | 0.467 |
| -0.92 | 0.002 | 0.480 | 0.018 |
| -1.44 | 0.485 | 0.004 | 0.012 |

Y point

| energy of the tight-binding calculation (eV) | $x'y'$ | $y'z'-z'x'$ | $y'z'+z'x'$ |
| --- | --- | --- | --- |
| 0.75 | 0.488 | 0.003 | 0.009 |
| 0.02 | 0.011 | 0.285 | 0.205 |
| -0.08 | 0.001 | 0.207 | 0.292 |
| -0.64 | 0.000 | 0.101 | 0.398 |
| -0.78 | 0.008 | 0.398 | 0.094 |
| -1.32 | 0.492 | 0.006 | 0.002 |

phase, and these spins form the singlet dimers as in spin-Peierls transitions of quasi-one-dimensional spin systems. For $Tl_2Ru_2O_7$, spins ($S=1$), which exist even above $T_c$, have been proposed to undergo the transition to a nonmagnetic state by the same mechanism as the Haldane gap formation well-known for quasi-one-dimensional spin systems. For these mechanisms, we cannot expect such a significant bond-length change as observed for the present system. This difference can be naturally understood, because $La_4Ru_2O_{10}$ and $Tl_2Ru_2O_7$ have the corner-sharing linkages of the $RuO_6$ octahedra, where the Ru-Ru bond lengths are much larger than that of $Li_2RuO_3$. The short Ru-Ru bond length or the edge-sharing linkage is the necessary condition to have the molecular-orbit formation.

The mechanism of the structural transition observed in $AlV_2O_4$[18,19] with the spinel structure is rather similar to that observed in $Li_2RuO_3$. In $AlV_2O_4$, V trimers seem to be formed by a molecular-orbit formation of vanadium 3$d$ $t_{2g}$ orbits at the transition temperature $T_c \sim 700$ K. In this case, the electrons which remain in the unpaired state gradually go to the nonmagnetic state with decreasing $T$, indicating that ground state has a spin gap. For this system, it is not clear whether the transition at 700 K is continuous or discontinuous.

Now, we have shown that the molecular-orbit formation is the primary driving force of the structural transition of $Li_2RuO_3$. We note that the transition seems to be second order. As far as we know, it is a first example of the continuous (or at least nearly continuous) structural transitions driven by the molecular orbit formation which takes place with varying temperature.

## 5. Summary

$Li_2RuO_3$ with the pseudo honeycomb structure exhibits a structural phase transition at $T_c \sim 540$ K, where two of the six bond lengths between Ru-Ru pairs within a hexagon exhibits a significant reduction upon the structural transition. The electrical resistivity $\rho$ exhibits an anomalous increase at $T_c$ with decreasing $T$, and the magnetic susceptibility $\chi$ also shows a sharp decrease. The density functional calculations for the system have been performed with GGA by the WIEN2k package. The calculated DOS data indicate that the metallic state is realized in the high-$T$ phase, and an energy gap exists at low-$T$. We have carried out the tight-binding fits to the dispersion curves obtained by GGA. In the high-$T$ phase, the characters of the obtained bands do not indicate any evidence for the characteristic weight distribution of the Ru 4$d$ $t_{2g}$ orbits. In contrast, below $T_c$, the electron bands split into the bonding ($\sigma$, $\pi$ and $\delta$) and antibonding ($\sigma^*$, $\pi^*$ and $\delta^*$) ones due to the very drastic distortion to form the Ru-Ru pairs. These results confirm our previous proposal that the lattice distortion forming Ru-Ru pairs is driven by the molecular-orbit formation. It is, as far as we know, the first or a quite rare example of the structural transition.

## Acknowledgements

The authors thank the condensed-matter theory group of department of physics of Nagoya University, especially Drs. M. Tsuchiizu and A. Kobayashi for computer analyses. This work is supported by Grants-in-Aid for Scientific Research from the Japan Society for the Promotion of Science (JSPS) and by Grants-in-Aid on priority areas from the Ministry of Education, Culture, Sports, Science and Technology.